\documentclass[cits,A4]{PoS}

\newcommand\ep{\varepsilon}

\newcommand\beq{\begin{equation}}
\newcommand\eeq{\end{equation}}
\newcommand\bea{\begin{eqnarray}}
\newcommand\eea{\end{eqnarray}}

\newcommand{\Ahathat}{\hat{\hspace*{-0.3mm}\hat{A}}}

\newcommand{\BLB}{\left[ \phantom{\rule{0.1mm}{0.56cm}} \! \right.}

\newcommand{\BRB}{\left. \! \phantom{\rule{0.1mm}{0.56cm}} \right] }

\title{{\tiny DESY 11--227,~~D0--TH 11/27,~~SFB/CPP-11-70,~~LPN11-67.}\\
Two-loop operator matrix elements for massive fermionic local twist-2 operators in QED}

\ShortTitle{Two-loop massive fermionic OME's in QED}

\author{J. Bl\"umlein,$^a$ \speaker{A. De Freitas}$^{ab}$ and W.L. van Neerven$^c$\thanks{Deceased}\\
        \llap{$^a$} DESY, Zeuthen, Platanenalle 6, D-15738 Zeuthen, Germany.\\
        \llap{$^b$} Departamento de F\'{i}sica, Universidad Sim\'{o}n Bol\'{i}var. Caracas 1080-A, Venezuela.\\
        \llap{$^c$} Institut-Lorentz, Universiteit Leiden, P.O. Box 9506, 2300 HA Leiden, The Netherlands.\\
        E-mail: \email{Johannes.Bluemlein@desy.de}, \email{abilio.de.freitas@desy.de} }

\abstract{We describe the calculation of the two--loop massive operator matrix
elements with massive external fermions in QED. We investigate the factorization
of the $O(\alpha^2)$ initial state corrections to $e^+e^-$ annihilation into a virtual
boson for large cms energies $s \gg m_e^2$ into massive operator matrix elements and 
the massless Wilson coefficients of the Drell-Yan process adapting the color coefficients
to the case of QED, as proposed by Berends et. al. in Ref. [1]. Our calculations show
explicitly that the representation proposed in Ref. [1] works at one-loop order and up 
to terms linear in $\ln(s/m^2_e)$ at two-loop order. However, the two-loop constant part 
contains a few structural terms, which have not been obtained in previous direct calculations.}

\FullConference{ 10th International Symposium on Radiative Corrections 
(Applications of Quantum Field Theory to Phenomenology) - Radcor2011\\
September 26-30, 2011\\
Mamallapuram, India}

\begin{document}


Ever since the operator product expansion formalism was applied to the analysis of 
deep inelastic scattering (DIS), there has been a lot of interest in the calculation 
of massive operator matrix elements at higher order in perturbation theory. 
Heavy flavor corrections to DIS structure functions are very important at small
values of the Bjorken variable $x$ (where they contribute on the level of 20--40$\%$), 
and can be calculated in the limit $Q^2 \gg m^2$ as a convolution of the corresponding massive 
operator matrix elements and the light flavor Wilson coefficients \cite{HEAV1}. Here $Q^2$ denotes
the virtuality of the gauge boson exchanged in DIS and $m$ is the mass of the heavy quark.

The scaling violations of the heavy flavor part in the structure functions $F_{2,L}(x,Q^2)$ 
are very different from those of the light flavor contributions, and their knowledge is very
important for precision measurements of $\Lambda_{\rm QCD}$ and the extraction of light parton 
densities.
A semi-analytic calculation of the heavy flavor contributions was done in Ref. \cite{Laenen} at 
next-to-leading order for the full kinematic range, and a fast numerical implementation in 
Mellin space was given in \cite{AleBlu}. Full analytic results in the 
asymptotic region $Q^2 \gg m^2$ for the structure function 
$F_2^{Q\bar{Q}}(x,Q^2)$ at $O(\alpha_s^2)$  were derived in Ref. \cite{HEAV1}, and recently 
recalculated in \cite{BBK}. In the same kinematic region, $F_L^{Q\bar{Q}}(x,Q^2)$ was 
obtained at $O(\alpha_s^3)$ in Ref. \cite{FreiBlu}. 

The $O(\alpha_s^3)$ massive operator matrix elements
are required in both cases and all but \linebreak $O((m^2/Q^2)^k)$, $k>1$
contributions are found. In this approximation, the structure function $F_2(x,Q^2)$ turns 
out to be very well described for $Q^2>20 \rm GeV^2$, while for $F_L(x,Q^2)$ this approximation 
only holds at large scales $Q^2 > 1000 \, \rm GeV^2$.

More recently, there has been considerable progress in the calculation of the $O(\alpha_s^3)$
heavy flavor contributions to the Wilson coefficients of the structure function $F_2(x,Q^2)$
and the massive gluonic operator matrix elements. First, in Ref. \cite{Gluonic} the $O(\alpha_s^2\ep)$
corrections to these matrix elements were given. These corrections are required to perform the
corresponding renormalization procedure at $O(\alpha_s^3)$. Later, the calculation of these 
contributions for a number of fixed moments $N = 2 ... 10 (12,14)$ at $O(\alpha_s^3)$ has also been achieved in 
\cite{MellinMomHF}, and 
$O(\alpha_s^3 n_f)$ contributions were given in Ref. \cite{Onf}. $O(\alpha_s^2)$ and $O(\alpha_s^3)$ 
heavy flavor contributions to transversity have also been obtained in \cite{HFTransversity}.

On the other hand, massive operator matrix elements, and in particular those
with a massive external fermion line, can also be applied to a different kind of
problem, namely, the calculation of initial state QED corrections of scattering
processes, such as $e^+e^-$ annihilation into a virtual gauge boson, using the 
renormalization group technique. A wealth of 
information about the Standard Model has been obtained in the past from 
electron--positron colliding beam experiments at different facilities around the world. 
In the future, projects like ILC \cite{ILC} and CLIC \cite{CLIC} are planned to put the 
Standard Model to even more decisive tests and to reveal new physics 
\cite{ILC:PHYS}. In this context high-luminosity machines which operate at a 
narrow energy regime as DAFNE \cite{DAFNE} and GIGA-Z \cite{ILC,ILC:PHYS} at 
the $Z$-peak will offer much higher precision on rare processes. 

The QED initial state radiation causes large corrections for various 
differential and integral scattering cross sections, depending on the 
sensitivity of the sub--system cross section with respect to kinematic
rescaling of variables and has to be known at sufficient precision.
Both for $e^+e^-$ annihilation at resonance peaks and the wings of resonances
at very high luminosities, the knowledge of the $O(\alpha^2)$ corrections
is mandatory to cope with the experimental precision.
While the $O(\alpha)$ corrections are known for a large amount of reactions, 
the corrections beyond the universal contributions $O((\alpha 
L)^k), 1 \leq k \leq 5$, see \cite{UNIV1,UNIV2}, to higher orders, were only 
calculated once at two-loop order in Ref.~\cite{BBN}. Besides the logarithmic 
orders $O(\alpha^2 L^2, \alpha^2 L)$ with $L = \ln(s/m_e^2)$ and $m_e$ the electron 
mass, the constant terms $O(\alpha^2)$ are of interest. 

The renormalization group technique allows to decompose the 
scattering cross section $\sigma(e^+ e^-$ 
$\rightarrow~V),~V = \gamma^*, Z^*$ 
into massive operator matrix elements and massless Wilson coefficients. The 
fermion mass effects are contained in the former, while the sub-system hard 
scattering cross sections are calculated for massless particles.
The corresponding massless Wilson coefficients 
are known from the literature \cite{DY1,DY2} for the Drell-Yan process.
In \cite{BBN} this method was used to derive all
terms up to ${\rm O}(\alpha^2L)$ in addition to the direct calculation. 

The differential scattering cross section can be written in the limit $s \gg m_e^2$
as a sum of three contributions \cite{BBN}:
\begin{equation}
\label{eq:XS}
\frac{d{\sigma}_{e^+e^-}}{ds'} = \frac{d{\sigma}_{e^+e^-}^{\rm I}}{ds'}
+ \frac{d{\sigma}_{e^+e^-}^{\rm II}}{ds'} + \frac{d{\sigma}_{e^+e^-}^{\rm III}}{ds'}~,
\end{equation}
where the labels I, II and III refer to the flavor non-singlet terms with a single fermion 
line, those with an additional closed fermion line, and the pure-singlet terms, respectively.
Here, $s'$ denotes the invariant mass of the virtual vector boson and $s$ the cms energy of the
process.
\begin{equation}
s'=xs, \quad 0 \leq x \leq 1.
\end{equation}

It is convenient to write the scattering cross section in Mellin space by applying the
integral transform
\begin{equation} 
\widehat{\frac{d\sigma}{ds'}}(N) = \int_0^1~dx~x^{N-1}~\frac{d\sigma}{ds'}(xs)~.
\end{equation}

Using the renormalization group method it can be shown that the three contributions 
in Eq. (\ref{eq:XS}) can be expressed as \cite{BBN,self}
\begin{eqnarray} 
\label{eqMA1a}
\widehat{
\frac{d\sigma_{e^+e^-}^{\rm I}}{ds'}} &=&
\frac{1}{s} \widehat{\sigma^{(0)}} 
   \Biggl\{
   1 + a_0 \left[ P_{ee}^{(0)}  {\bf L}
   +\tilde{\sigma}^{(0)}_{ee} + 2 \Gamma_{ee}^{(0)}\right] 
   + a_0^2\Biggl[
   \frac{1}{2} {P_{ee}^{(0)}}^2 {\bf L}^2
\nonumber \\ && \phantom{\frac{1}{s}}
   +\left( P_{ee}^{(1),{\rm I}} 
   + P_{ee}^{(0)} \left( \tilde{\sigma}_{ee}^{(0)} + 
   2 \Gamma_{ee}^{(0)}\right) \right) {\bf L}
   + 2 \Gamma_{ee}^{(1),{\rm I}} + 
\tilde{\sigma}_{ee}^{(1),{\rm I}} + 2 \Gamma_{ee}^{(0)} 
\tilde{\sigma}_{ee}^{(0)} 
+ {\Gamma_{ee}^{(0)}}^2 \Biggr] \Biggr\}~,
\\
\label{eqMA1b}
\widehat{\frac{d\sigma_{e^+e^-}^{\rm II}}{ds'}} &=&
\frac{1}{s} \widehat{\sigma^{(0)}} 
   a_0^2\Biggl\{- \frac{\beta_0}{2} P_{ee}^{(0)} {\bf L}^2
   +\Biggl[P_{ee}^{(1), {\rm II}} - \beta_0 \tilde{\sigma}_{ee}^{(0)} 
     \Biggr] {\bf L}
   + 2 \Gamma_{ee}^{(1),{\rm II}} + 
\tilde{\sigma}_{ee}^{(1),{\rm II}} \Biggr\}~,
\\
\label{eqMA1c}
\widehat{\frac{d\sigma_{e^+e^-}^{\rm III}}{ds'}} &=&
\frac{1}{s} \widehat{\sigma^{(0)}} 
   a_0^2\Biggl\{\frac{1}{4}
   P_{e \gamma}^{(0)}  P_{\gamma e}^{(0)} {\bf L}^2
   +\Biggl[P_{ee}^{(1),{\rm III}} + P_{\gamma e}^{(0)} 
\tilde{\sigma}_{e 
\gamma}^{(0)}
   + \Gamma_{\gamma e}^{(0)} P_{e \gamma}^{(0)} \Biggr] {\bf L}
   \nonumber\\
 & & \phantom{ \frac{1}{s} \widehat{\sigma^{(0)}} a_0^2}
   + 2 \Gamma_{ee}^{(1),{\rm III}} + \tilde{\sigma}_{ee}^{(1),{\rm III}}
+ 2 \tilde{\sigma}_{e \gamma}^{(0)} 
   \Gamma_{\gamma e}^{(0)}
\Biggr\}~,
\end{eqnarray} 
where ${\bf L}=\ln(s/m_e^2)+\ln(x)$, and $\widehat{\sigma^{(0)}}$ is the Born 
cross section. The quantities $P_{ij}^{(0)}$ and $\tilde{\sigma}_{ij}^{(0)}$
are the LO splitting functions and LO Wilson coefficients, respectively, while at NLO
they are denoted by  $P_{ee}^{(1)}$ and $\tilde{\sigma}_{ee}^{(1)}$. 
$\Gamma_{ij}^{(0)}$ and $\Gamma_{ee}^{(1)}$ are the one-loop and two-loop constant terms
of the massive operator matrix elements, respectively. The labels I, II and III appear in 
$P_{ee}^{(1)}$, $\tilde{\sigma}_{ee}^{(1)}$ and $\Gamma_{ee}^{(1)}$ corresponding to 
the three possible contributions. The constant $\beta_0$ is the first coefficient
in the expansion of the QED $\beta$-function.

The splitting functions up to $O(\alpha^2)$ are well known \cite{PijNLO}, and as we mentioned before,
so are the massless Wilson coefficients \cite{DY1,DY2}. Here we present the 
missing ingredient needed to complete the decomposition of the scattering cross section 
according to Eqs.~(\ref{eqMA1a}--\ref{eqMA1c}), namely, the $O(\alpha^2)$ massive operator 
matrix elements. Details on the calculation can be found in Ref. \cite{self}.

The bare operator matrix elements are given by
\begin{eqnarray}
\Ahathat _{ij}\left(\frac{{m}^2_e}{\mu^2},\ep,N\right) = \delta_{ij} + \sum_{k=1}^\infty
\hat{a}^k \Ahathat_{ij}^{(k)}\left(\frac{{m}^2_e}{\mu^2},\ep,N\right)~,
\end{eqnarray}
where $\hat{a}$ is the unrenormalized coupling constant. The double hat means that the
quantity is completely unrenormalized. The complete renormalization procedure includes
charge renormalization, wave function renormalization and the renormalization of the 
composite operators. We need the inclusion of some counterterm diagrams in the case of
process I. The electron mass is renormalized on-shell, $p^2=m^2_e$, with $p$ the 
momentum of the external fermion, which means that no collinear singularities will
appear.

\begin{figure}
\begin{center}
\includegraphics[scale=.8]{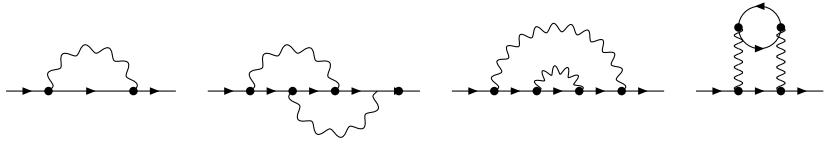}
\end{center} 
\caption{The self-energy diagrams. \label{SelfEnergy}}
\end{figure}

The wave function renormalization was performed using the Z-factors coming from the fermion
self-energy, see Fig. \ref{SelfEnergy}. Since we have a massive fermion in the external legs,
the coupling constant was first obtained in the {\sf MOM}-scheme, after which we transform to
the $\overline{\rm MS}$ scheme using
\begin{eqnarray}
\label{eq:aTR}
a^{\rm MOM}
= a^{\overline{\rm MS}} +\frac{4}{3} \ln\left(\frac{m_e^2}{\mu^2}\right) {a^{\overline{\rm MS}}}^2 
+ O\left({a^{\overline{\rm MS}}}^3\right)~.
\end{eqnarray}

We perform the calculation in $D=4+\varepsilon$ dimensions.
It can be shown that after wave function and charge renormalization, keeping the charge in
the {\sf MOM}-scheme, the two-loop OMEs, denoted by a single hat, are given by
\begin{eqnarray}
\label{unA2I}
\hat{A}_{ee}^{\rm I} &=& a^{\sf MOM}  S_\varepsilon \left(\frac{m^2_e}{\mu^2}\right)^{\ep/2}
\left[-\frac{1}{\ep} P_{ee}^{(0)} + \Gamma_{ee}^{(0)} + \ep \overline{\Gamma}^{(0)}_{ee}\right] 
\nonumber\\ &&
+ {a^{\sf MOM}}^2  S_\varepsilon^2 
                       \left(\frac{m^2_e}{\mu^2}\right)^{\varepsilon} 
\left\{\frac{1}{2 \varepsilon^2} P_{ee}^{(0)} \otimes P_{ee}^{(0)} 
- \frac{1}{2 \varepsilon} \left[P_{ee}^{(1), \rm I} + 2 \Gamma_{ee}^{(0)}
\otimes P_{ee}^{(0)}\right] + \hat{\Gamma}_{ee}^{(1), \rm I}\right\}~,
\\
\label{unA2II}
\hat{A}_{ee}^{\rm II} &=& {a^{\sf MOM}}^2  S_\varepsilon^2 
                       \left(\frac{m^2_e}{\mu^2}\right)^{\varepsilon} 
\left\{\frac{1}{ 2\varepsilon^2}  2\beta_0 P_{ee}^{(0)} 
- \frac{1}{2\varepsilon}  \left[ 
P_{ee}^{(1), \rm II} + 4 \beta_0 \Gamma_{ee}^{(0)} \right] + 
\hat{\Gamma}_{ee}^{(1), \rm II}\right\}~,
\\
\label{unA2III}
\hat{A}_{ee}^{\rm III} &=& {a^{\sf MOM}}^2  S_\varepsilon^2 
                       \left(\frac{m^2_e}{\mu^2}\right)^{\varepsilon} 
                       \left\{\frac{1}{2 \varepsilon^2} P_{e 
\gamma}^{(0)} \otimes P_{\gamma e}^{(0)} - \frac{1}{2 \varepsilon} 
\left[ P_{ee}^{(1), \rm III} + 2 \Gamma_{\gamma e}^{(0)} \otimes 
P_{e \gamma}^{(0)} \right]  
+ \hat{\Gamma}_{ee}^{(1), \rm III} \right\}~.
\end{eqnarray}

For the final renormalization step, i.e., the renormalization of the composite operators
we just need to include the corresponding inverse Z-factors, $Z^{(1)}_{ij}$, after which
the completely renormalized OMEs are given by

\begin{eqnarray}
A_{ij}^{\sf MOM} = \delta_{ij} 
                   + a^{\sf MOM} \left[\hat{A}_{ij}^{(1)} + Z_{i,j}^{-1,(1)} \right]
                   + {a^{\sf MOM}}^2 
                     \left[\hat{A}_{ij}^{(2)} + Z_{i,j}^{-1,(2)} 
+ Z_{i,j}^{-1,(1)} \hat{A}_{ij}^{(1)}\right] +O({a^{\sf MOM}}^3)~.
\end{eqnarray}

\begin{figure}
\centering
\begin{tabular}{ccccc}
\includegraphics[scale=0.55]{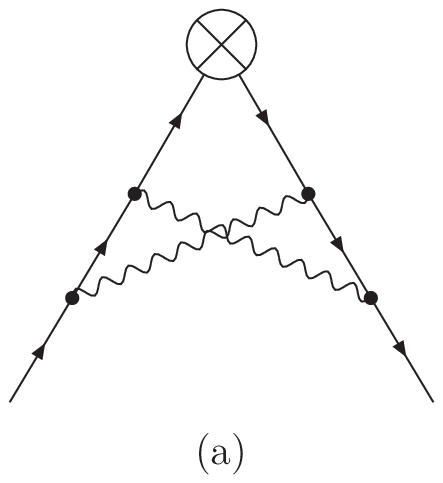} &
\includegraphics[scale=0.55]{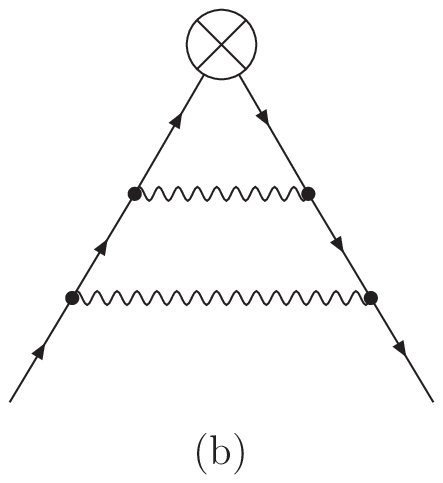} &
\includegraphics[scale=0.55]{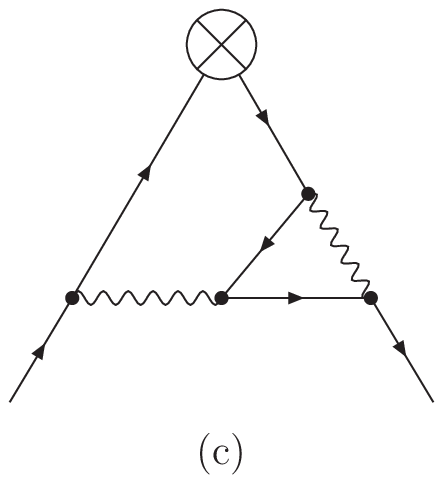} &
\includegraphics[scale=0.55]{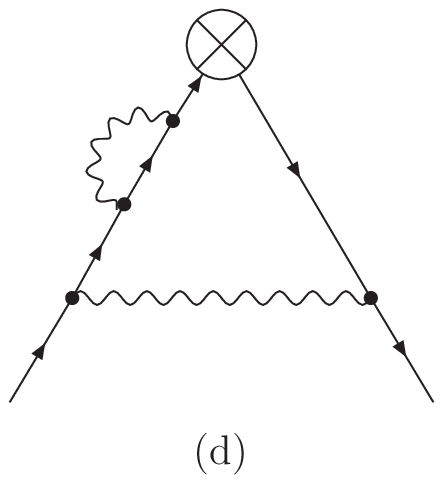}  &
\includegraphics[scale=0.55]{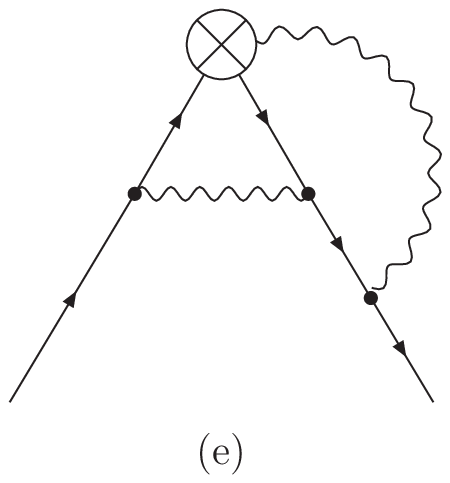} \\
\includegraphics[scale=0.55]{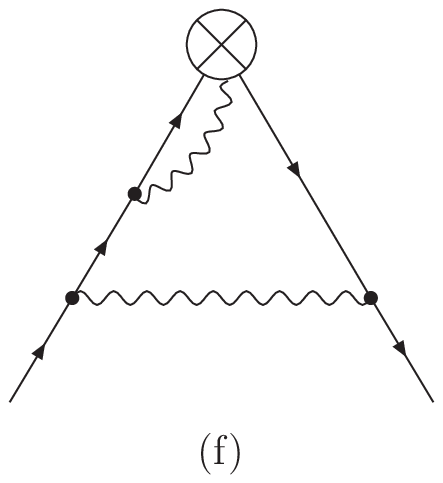} &
\includegraphics[scale=0.55]{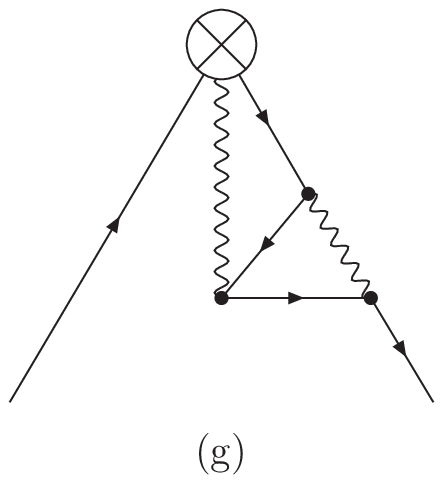} &
\includegraphics[scale=0.55]{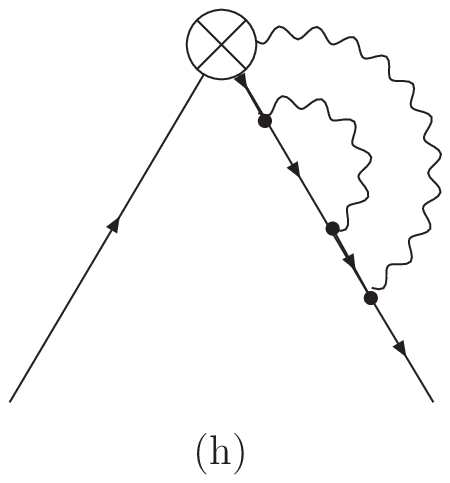} &
\includegraphics[scale=0.55]{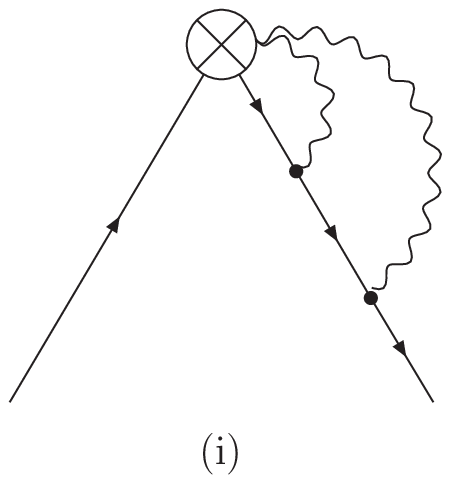} &
\includegraphics[scale=0.55]{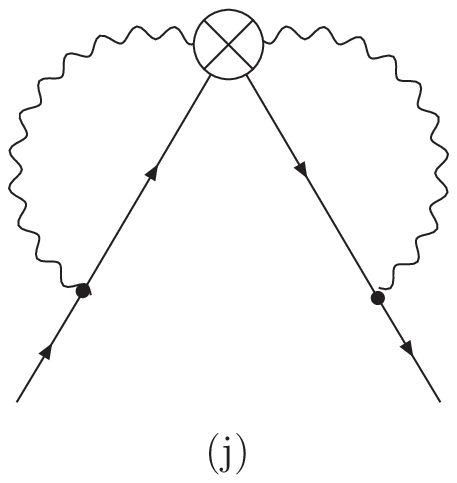}
\end{tabular}
\caption{
Feynman diagrams for the calculation of the massive two-loop operator
matrix elements $A^{(2), {\rm I}}_{ee}$. \label{DiagramsI}}
\end{figure}

\begin{figure}
\centering
\begin{tabular}{rl}
\includegraphics[scale=0.7]{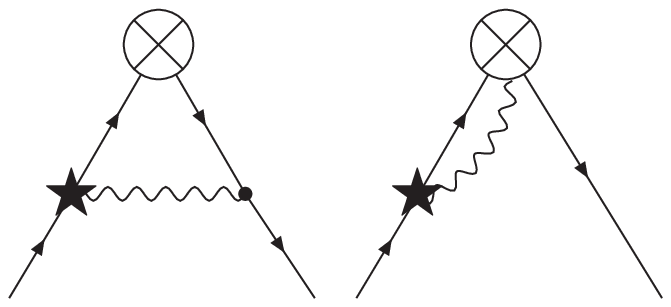} &
\includegraphics[scale=0.7]{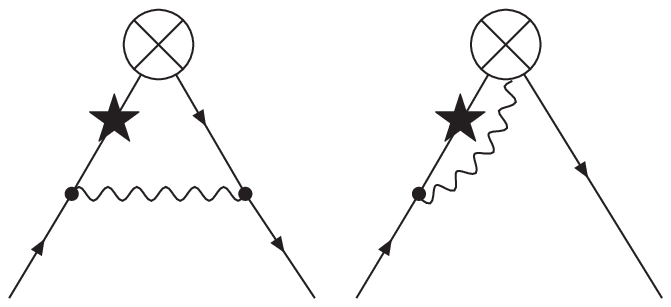} 
\end{tabular}
\caption{Counterterm diagrams. The black stars represent the counterterm vertices.
\label{CounterTermDiags}}
\end{figure}

The Feynman diagrams required for process I are shown in Fig. \ref{DiagramsI}.
As it was mentioned before, in order to renormalize the corresponding OMEs for
this process, we need to include also the counterterm diagrams shown in Fig.
\ref{CounterTermDiags}, where the black stars represent the counterterm 
vertices in QED. The diagrams are calculated using the standard Feynman rules
for the operator insertions, and projecting the resulting numerators in the
integrands with the factor $(\not \! \! p + m)/4$, after which we take the trace.
The program {\tt FORM}  \cite{math-ph/0010025}
was used to decompose the diagrams as a linear combination
of integrals with different powers of propagators, which arise after canceling as many
terms as possible in the numerator against the propagators. The most complicated
integrals appearing in the expressions are those where all five propagators are present. 
These can be computed using integration by parts identities to express them in terms of
4-propagator integrals. The resulting integrals were checked by several means, including
their representation in terms of Mellin-Barnes integrals, cf. \cite{self}.

%
%

The constant contributions to process I in Eq. (\ref{unA2I}) is given in $x$-space by 
\cite{self}
\begin{eqnarray}
\label{eq:GA1}
\hat{\Gamma}^{(1),\rm I}_{ee} &=&  
\frac{1+3 x^2}{1-x} \left[ 6 \zeta_2 \ln(x) -8 \ln(x) {\rm Li}_2(1-x)-4 \ln^2(x) \ln(1-x) \right]
+\left( {122 \over 3} x+22+{32 \over 1-x} \right) \zeta_2
\nonumber \\ &&
+16 \frac{1+x^2}{1-x} \left[ 2 {\rm Li}_3(-x)-\ln(x) {\rm Li}_2(-x) \right]
+{80 \over 3 (1-x)}
+56 (1+x) \zeta_2 \ln(1-x)
+\left( {22 \over 3} x+32
\right.
\nonumber \\ &&
\left.
+{64 \over 3 (1-x)^2}-{51 \over 1-x}
-{16 \over 3 (1-x)^3} \right) \ln^2(x)
-(92+20 x) \ln^2(1-x)
+\left( {178 \over 3}+{64 \over 3 (1-x)^2}
\right.
\nonumber \\ &&
\left.
-36 x -{140 \over 3 (1-x)}-{48 \over 1+x} \right) \ln(x) 
-\frac{1}{3} (1+x) \ln^3(x)
+4 \frac{x^2-8 x-6}{1-x} \ln(x) \ln(1-x) 
\nonumber \\ &&
-2 \frac{1+17 x^2}{1-x} \ln(x) \ln^2(1-x)
-\frac{112}{3} (1+x) \ln^3(1-x)
 +32 \frac{1+x}{1-x} \left[ \ln(x) \ln(1+x)+{\rm Li}_2(-x) \right]
\nonumber \\ &&
-22 x-{62 \over 3}
-4 \frac{13 x^2+9}{1-x} {\rm S}_{1,2}(1-x)
+4 \frac{5-11 x^2}{1-x} \left[ \ln(1-x) {\rm Li}_2(1-x)
-{\rm Li}_3(1-x)-2 \zeta_3 \right]
\nonumber \\ &&
+\frac{4 (16 x^2-10 x-27)}{3 (1-x)} {\rm Li}_2(1-x)
+14 (x-2) \ln(1-x)
+\left( 16-52 \zeta_2+128 \zeta_3 \right) {\cal D}_0(x)
\nonumber \\ &&
+\left( 8-112 \zeta_2 \right) {\cal D}_1(x)
+120 {\cal D}_2(x)
+{224 \over 3} {\cal D}_3(x)
+\left[ {433 \over 8}+58 \zeta_3
+\left( {37 \over 2}-48 \ln(2) \right) \zeta_2
\right.
\nonumber \\ &&
\left.
-{67 \over 45} \pi^4 \right] \delta(1-x)
+(-1)^n \left\{
\frac{4 (x^2+10 x-3)}{3 (1+x)} \left( \zeta_2+2 {\rm Li}_2(-x)+2 \ln(x) \ln(1+x) \right)
\right.
\nonumber \\ &&
+\frac{2 (1-x) (45 x^2+74 x+45)}{3 (1+x)^2}
+\frac{2 (9+12 x+30 x^2-20 x^3-15 x^4)}{3 (1+x)^3} \ln(x) 
\nonumber \\ &&
+\frac{1+x^2}{1+x} \BLB
8 \zeta_2 \ln(x)-24 \zeta_2 \ln(1+x)+36 \zeta_3
-\frac{2}{3} \ln^3(x)
+40 {\rm Li}_3(-x)
-4 \ln^2(x) \ln(1+x)
\nonumber \\ &&
-24 \ln(x) \ln^2(1+x)
-24 \ln(x) {\rm Li}_2(-x)
-48 \ln(1+x) {\rm Li}_2(-x)
-8 \ln(x) {\rm Li}_2(1-x)
\nonumber \\ &&
-16 {\rm S}_{1,2}(1-x)
-48 {\rm S}_{1,2}(-x)
\BRB
-\frac{16 (x^4+12 x^3+12 x^2+8 x+3)}{3 (1+x)^3} {\rm Li}_2(1-x)
\nonumber \\ &&
\left.
 +4 x \frac{1-x-5 x^2+x^3}{(1+x)^3} \ln^2(x) 
\right\}~,
\end{eqnarray}
where ${\rm Li}_n(x)$ and ${\rm S}_{n,p}(x)$ are the well known polylogarithm and Nielsen
functions, and 
\[
{\cal D}_k(x)=\left( \frac{\ln^k(1-x)}{1-x} \right)_+.
\]

\begin{figure}
\begin{center}
\begin{minipage}[c]{0.5\linewidth}
\centering
\includegraphics[scale=0.55]{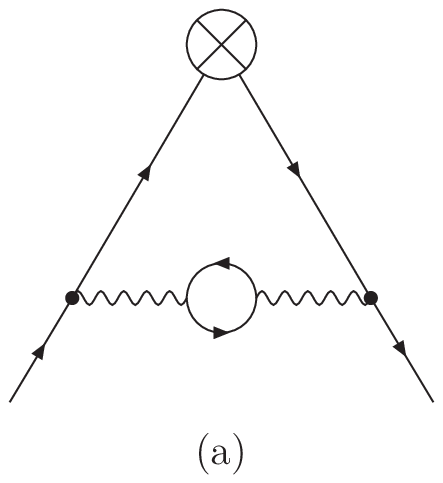} 
\end{minipage}
\hspace{-37mm}
\begin{minipage}[c]{0.5\linewidth}
\centering
\includegraphics[scale=0.55]{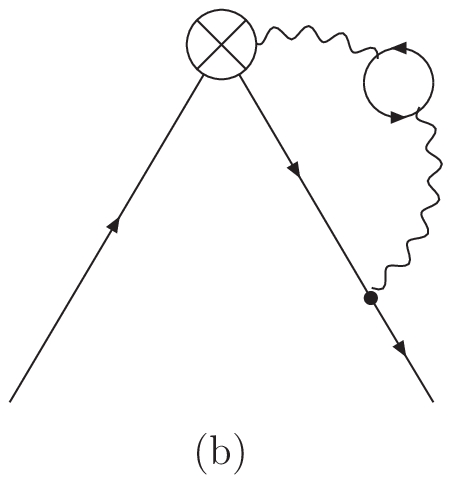}
\end{minipage}
\caption{
Feynman diagrams for the calculation of the massive two-loop operator
matrix elements $A^{(2), {\rm II}}_{ee}$. \label{DiagramsII}}
\end{center}
\end{figure}

For process II we need the diagrams in Fig. \ref{DiagramsII}. These are relatively
easy to compute, since they involve a one-loop insertion. The result for the constant
term of process II in Eq. (\ref{unA2II}) is given by \cite{self}
%
%
%
\begin{eqnarray}
\label{eq:GA2}
\hat{\Gamma}^{(1),\rm II}_{ee} &=& 
{76 \over 27} x-{572 \over 27}
-\left( 12 x+{4 \over 3}+{8 \over 1-x}
+{32 \over 9 (1-x)^2}-{160 \over 9 (1-x)^3}+{64 \over 9 (1-x)^4} \right) \ln(x)
+{128 \over 9 (1-x)^2}
\nonumber \\ &&
+{80 \over 27 (1-x)}
-{64 \over 9 (1-x)^3}
-{2 (1+x^2) \over 3 (1-x)} \ln^2(x)
+{16 \over 3} (1+x) \left( \ln(1-x)+\ln^2(1-x)+{1 \over 4} \zeta_2 \right)
\nonumber \\ &&
+\left( {224 \over 27} -{8 \over 3} \zeta_2 \right) {\cal D}_0(x)
-{32 \over 3} \left( {\cal D}_1(x)+{\cal D}_2(x) \right)
+\left( {8 \over 3} \zeta_3-10 \zeta_2
+{1411 \over 162} \right) \delta(1-x).
\end{eqnarray}

\begin{figure}
\begin{center}
\includegraphics[scale=0.6]{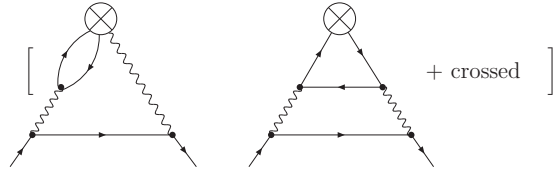}
\end{center}
\caption{
Feynman diagrams for the calculation of the massive two-loop operator
matrix elements $A^{(2),{\rm III}}_{ee}$. \label{DiagramsIII}}
\end{figure}

In the case of process III, we need the pure singlet diagrams shown in 
Fig.~\ref{DiagramsIII}. These can be calculated using the corresponding
one-loop operator insertions, which were given in Ref. \cite{KleinPHDThesis}. The
result for the constant term appearing in Eq. (\ref{unA2III}) is \cite{self}
%
%
%
\begin{eqnarray}
\label{eq:GA3}
\hat{\Gamma}^{(1),\rm III}_{ee} &=&
{2 \over x} (1-x) (4 x^2+13 x+4) \zeta_2
+{1 \over 3 x} (8 x^3+135 x^2+75 x+32) \ln^2(x)
+\left[ {304 \over 9 x}-{80 \over 9} x^2-{32 \over 3} x
\right.
\nonumber \\ &&
\left.
+108-{32 \over 1+x}-{64 (1+2 x) \over 3 (1+x)^3} \right] \ln(x)
+50
-{224 \over 27} x^2
-{182 \over 3} x
-{32 \over 1+x}
+{64 \over 3 (1+x)^2}
+{800 \over 27 x}
\nonumber \\ &&
+16 {1-x \over 3 x} (x^2+4 x+1) \left[ 2 \ln(x) \ln(1+x)
-{\rm Li}_2(1-x)+2 {\rm Li}_2(-x) \right]
+(1+x) \BLB
4 \zeta_2 \ln(x) 
\nonumber \\ &&
+{14 \over 3} \ln^3(x)
-32 \ln(x) {\rm Li}_2(-x)
-16 \ln(x) {\rm Li}_2(x)
+64 {\rm Li}_3(-x)
+32 {\rm Li}_3(x)
+16 \zeta_3
\BRB~. 
\end{eqnarray}

Our calculations show explicitly that the operator matrix elements satisfy the relations given in 
Eqs.~(\ref{unA2I}--\ref{unA2III}) for the pole terms, which automatically guarantees  that the 
decomposition given in Eqs.~(\ref{eqMA1a}--\ref{eqMA1c}) holds for the logarithmic terms, as found in Ref. \cite{BBN}. 
Furthermore, we have verified that the first moment of the operator matrix elements vanishes,
including the constant terms given in Eqs.~(\ref{eq:GA1}-\ref{eq:GA3}), 
so they obey fermion  number conservation as they should. This provides a highly 
non-trivial check of our results. However, when we assemble the final result for the constant
terms in Eqs.~(\ref{eqMA1a}--\ref{eqMA1c}) by including the massless Wilson coefficients, we observe 
that, although many of terms appearing in Ref. \cite{BBN} appear in our results, 
a few structural terms, such as terms proportional to $1/(1-x)^3$ and $\ln(x)/(1-x)^4$, also appear,
which were not present in the result given in \cite{BBN}. This result appears in contrast to the case
of massless external fermions and boson lines, where the corresponding cross sections have been
shown to factorize, including the constant terms, as can be seen in Refs. \cite{HEAV1,Laenen,BBK,FreiBlu,HQCD}. 
The issue requires further investigation in order to elucidate the reasons for this.

\end{document}